# LINEAR REGRESSION EVALUATION OF SEARCH ENGINE AUTOMATIC SEARCH PERFORMANCE BASED ON HADOOP AND R

Hong Xiong

University of California – Los Angeles, Los Angeles, CA, USA

*ABSTRACT*

*The automatic search performance of search engines has become an essential part of measuring the difference in user experience. An efficient automatic search system can significantly improve the performance of search engines and increase user traffic. Hadoop has strong data integration and analysis capabilities, while R has excellent statistical capabilities in linear regression. This article will propose a linear regression based on Hadoop and R to quantify the efficiency of the automatic retrieval system. We use R's functional properties to transform the user's search results upon linear correlations. In this way, the final output results have multiple display forms instead of web page preview interfaces. This article provides feasible solutions to the drawbacks of current search engine algorithms lacking once or twice search accuracies and multiple types of search results. We can conduct personalized regression analysis for user's needs with public datasets and optimize resources integration for most relevant information.*

*KEYWORDS*

*Hadoop, R, search engines, linear regression, machine learning.*

## 1. INTRODUCTION

With the rapid development of the Internet, the Internet has gradually penetrated all aspects of users' lives and work. People can search and obtain the information they want through the information system platform [1]. In traditional information retrieval systems, people tend to focus on retrieval techniques, algorithms, and how to help users better provide information that matches keywords. However, the background and purpose of the user search are different. Traditional information retrieval systems cannot meet the requirements of users. With the emergence of social search platforms such as social media and social question and answer systems, users are no longer limited to the "human-machine" interaction model. With social services such as making friends, cooperating, sharing, communicating, and publishing content, users can quickly and accurately find information to meet their needs [2].

The search engine is a necessary function for the convenience of the source users to use the website to construct the website. It is also a useful tool for studying the website users' behaviour. New Competitiveness believes that efficient site search can allow users to find target information quickly and accurately, thereby more effectively promoting the sales of products/services. Through in-depth analysis of website visitors' search behaviour, it is helpful for further development of a more effective network marketing strategy. Therefore, for essential information websites with rich content and online sales websites with rich product lines, it is far from enough to provide general full-text search. It is necessary to develop advanced search functions that can achieve personalized needs and reflect on the crucial aspects of the website's network marketing function.

The search engine has become an indispensable tool of the Internet, which can help people find the content and information they want more quickly, improve the efficiency of doing things, and efficiently use Internet resources.

However, users now generally have secondary searches when they use search engines. This phenomenon is the fundamental basis of the writing of this paper. Moreover, many secondary searches are further attributive restrictions on nouns, showing that the search results that users need are no longer just the abbreviated content of the webpage but also require the participation of rich elements. However, due to our lack of professional knowledge and understanding of search engines, we cannot further analyse the underlying causes and propose practical solutions. We can only hypothesize and demonstrate our conjectures. This work needs to be further improved, and we look forward to seeing perfect theoretical research results from other scholars. Nevertheless, this paper provides a feasible algorithm combined with Hadoop and R for optimizing resource integration of search engines, along with a program frame for the realization of this algorithm.

In the second section, this paper will discuss related works about optimizations of search engine algorithms and their drawbacks. In the third section, we will discuss the properties of R and Hadoop separately and their integration basis. In the fourth section, we will propose a R-based Hadoop vision for algorithm optimization, reason of choosing linear regression, market value of this proposal, program frame and related experiments. In the final section, we will summarize all the assumptions and limitations of our proposal and analyse the next step of our research.

## 2. RELATED WORKS

Performance evaluation has always been one of the core issues of network information retrieval research. Traditional evaluation methods require a lot of human resources and material resources. Based on user behaviour analysis, a method for automatically evaluating search engine performance is proposed [3]. The navigation type queries the test set and automatically annotates the standard answers corresponding to the query [4]. Experimental results show that this method can achieve a basic performance. This consistent evaluation effect dramatically reduces the workforce and material resources required for evaluation and speeds up the evaluation feedback cycle.

The retrieval system's evaluation problem has always been one of the core problems in information retrieval research. Saracevic pointed out: "Evaluation problem is in such an important position in the research and development process of information retrieval that any new method and their evaluation. The way is integrated." Kent first proposed the precision rate-recall rate information retrieval evaluation framework. Subsequently, research institutions affiliated with the US government began to strongly support research on retrieval evaluation and the United Kingdom's Cranfield project in the late 1950s. The evaluation plan based on query sample sets, standard answer sets, and corpus established in the mid-1960s truly made information retrieval an empirical discipline and thus established the core of evaluation in information retrieval research. Status and its evaluation framework are generally called the Cranfield-like approach (A Cranfield-like approach) [5].

The Cranfield method points out that the evaluation of an information retrieval system should consist of the following links:

First, determine the set of query samples, extract a part of the query samples that best represent the user's information needs, and build a set of appropriate scale.

Second, focus on the query samples Set, find the corresponding answer in the corpus that the retrieval system needs to retrieve, that is, mark the standard answer set.

Finally, enter the query sample set and corpus into the retrieval system.

The system feeds back the search results and then uses the search evaluation index to evaluate the search results' closeness and the standard answer. It gives the final evaluation results expressed in numerical values.

Cranfield method has been widely used in most information retrieval system evaluation work, including search engines. TREC (Text Information Retrieval Conference) jointly organized by the Defense Advanced Research Projects Agency (DARPA) and the National Institute of Standards and Technology (NIST) has been organizing information retrieval evaluation and technical exchange forums based on this method. In addition to TREC, some search evaluation forums based on the Cranfield method designed for different languages have begun to try and operate, such as the NTCIR (NACSIS Test Collection for IR Systems) program and the IREX (Information Retrieval and Extraction Exercise) program [6].

With the continuous development of the World Wide Web and the increase in the amount of information on the Internet, how to evaluate the performance of network information retrieval systems has gradually become a hot topic in the evaluation of information retrieval in recent years. The Cranfield method has encountered tremendous obstacles when evaluating this aspect. The difficulty is mainly reflected in the standard answer labelling for the query sample set. According to Voorhees's estimation, it takes nine reviewers a month to label a specific query sample's standard answer on a corpus of 8 million documents. Although Voorhees proposed labelling methods such as Pooling to relieve labelling pressure, it is still challenging to label answers to massive network documents. Such as TREC massive scale retrieval task (Terabyte Track). Generally, it takes more than ten taggers 2-3 months to tag about dozens of query samples and corpora.

According to the scale, it is only about 10 million documents. Considering that the index pages involved in current search engines are more than several billion pages (Yahoo! reports 19.2 billion pages, and Sougou's claimed index in Chinese is also more than 10 billion), the network information retrieval system is carried out by manually marking answers. The evaluation will be a labour-consuming and time-consuming process. Due to the need for search engine algorithm improvement, operation, and maintenance, the retrieval effect evaluation feedback time needs to be shortened as much as possible. Therefore, improving the automation level of search engine performance evaluation is a hot spot in the current retrieval system evaluation research.

## 3. HADOOP & R

### 3.1. Hadoop

Hadoop is a distributed system infrastructure developed by the Apache Foundation [7]. Users can develop distributed programs without understanding the underlying details of distributed and make full use of the power of clusters for high-speed computing along with storage. Hadoop implements a distributed file system (Hadoop Distributed File System), one of which is HDFS [8].

HDFS has the characteristics of high fault tolerance and is designed to be deployed on low-cost hardware. It provides high throughput to access application data, and it is suitable for large dataset applications. HDFS relaxes POSIX requirements and can access data in the file system in the form of streaming access. The core design of the Hadoop framework is HDFS and MapReduce. HDFS provides storage for massive amounts of data, while MapReduce provides calculations for massive amounts of data [9].

### 3.2. R

R provides a wide variety of statistical (linear and nonlinear modelling, classical statistical tests, time-series analysis, classification, clustering) and graphical techniques. Moreover, it is highly extensible. The S language is often the vehicle of choice for research in statistical methodology, and R provides an Open Source route to participation in that activity [10]. Also, R is now the most widely used statistical software in academic science and it is rapidly expanding into other fields such as finance. R is almost limitlessly flexible and powerful [11].

One of R's strengths is the ease with which well-designed publication-quality plots can be produced, including mathematical symbols and formulae where needed. Great care has been taken over the defaults for the minor design choices in graphics, but the user retains full control [12].

R is an integrated suite of software facilities for data manipulation, calculation, and graphical display [13]. It includes an effective data handling and storage facility.

I. a suite of operators for calculations on arrays, in particular matrices,
II. an extensive, coherent, integrated collection of intermediate tools for data analysis,
III. graphical facilities for data analysis and display either on-screen or on hardcopy, and
IV. a well-developed, simple, and effective programming language, including conditionals, loops, user-defined recursive functions, and input and output facilities.

The term "environment" is intended to characterize it as a thoroughly planned and coherent system, rather than an incremental accretion of particular and inflexible tools, as is frequently the case with other data analysis software.

Like S, R is designed around an actual computer language, and it allows users to add additional functionality by defining new functions. Much of the system is itself written in the R dialect of S, making it easy for users to follow the algorithmic choices made. For computationally intensive tasks, C, C++, and Fortran code can be linked and called at run time. Advanced users can write C code to manipulate R objects directly [14].

Many users think of R as a statistics system [15]. We prefer to think of it as an environment within which statistical techniques are implemented. R can be extended (easily) via packages. There are about eight packages supplied with the R distribution, and many more are available through the CRAN family of Internet sites covering an extensive range of modern statistics.

For hardware reasons (disk space, CPU performance) there is currently no search facility at the R master webserver itself. However, due to the highly active R user community (without which R would not be what it is today) there are other possibilities to search in R web pages and mail archives:

An R site search is provided by Jonathan Baron at the University of Pennsylvania, United States. This engine lets you search help files, manuals, and mailing list archives [16].

Rseek is provided by Sasha Goodman at Stanford University. This engine lets you search several R-related sites and can easily be added to the toolbar of popular browsers [17].

The Nabble R Forum is an innovative search engine for R messages. As it has been misused for spam injection, it is nowadays severely filtered. In addition, its gateway to R-help is sometimes not bidirectional, so we do not recommend it for posting (rather at most for browsing) [18].

### 3.3. R and Hadoop Integration Base

R is a complete data processing, calculation, and drawing software system. The idea of R is it can provide some integrated statistical tools, but a more considerable amount is that it provides various mathematical calculations and statistical calculation functions so that users can flexibly analyse data and even create new ones that meet their needs [19].

Hadoop is a framework for distributed data and computing. It is good at storing large amounts of semi-structured data sets. Data can be stored randomly, so the failure of a disk will not cause data loss. Hadoop is also incredibly good at distributed computing-quickly processing large data sets across multiple machines [20].

Hadoop can be widely used in big data processing applications thanks to its natural advantages in data extraction, transformation, and loading (ETL). Hadoop has distributed architecture that puts the big data processing engine as close to the storage as possible, which is relatively suitable for batch processing operations such as ETL. The batch processing results of similar operations can go directly to storage. The MapReduce function of Hadoop realizes the fragmentation of a single task. It sends the fragmented task (Map) to multiple nodes and then loads (Reduce) into the data warehouse in the form of a single data set. When users search for information, do they only need a web-linked display, or do they need multimedia materials and resources such as pictures, videos, and audio-visual [21]?

## 4. R BASED HADOOP

For customers' keywords, Hadoop can respond quickly to the attached resources, but it cannot provide rich content and forms. R can compensate for this weakness. This issue is the form we want to explore today. It is possible to use R's functional computing capabilities based on Hadoop to quickly mobilize various forms of network resources to provide users with various high-value information.

In the global search, it is the display of web links. It pushes diversified information such as pictures and videos for users to choose personalized search, personalized settings, personalized data analysis, and personalized data output. Therefore, we might need to conduct forward-looking questions and answers on customer search requirements in advance, understand the main search requirements areas or directions of customers and reduce pushes in other areas.

The current search engines are all searched by the Hadoop algorithm. Now we will find out whether Hadoop can allow users to search for the desired results only once by using the search user usage of some search engines.

### 4.1. Linear Regression in Data Processing

In this paper, we choose linear regression as main method for the following reasons:

I. The linear regression has high speed in model-building and lack of overly complex calculation to minimize overfitting issues. The volatility of users' data requires a highly up-to-date analysis tool to optimize the present value, which can be satisfactorily handled by high speed of linear regression.

II. Linear regression provides coefficients of each variable for further explanation and analysis, which helps the researchers to interpret and conduct experiments upon each single variable. This interpretability cannot be matched with more complex tools from machine learning and deep learning.

III. Through non-linear transformations and generalized linear model, the linear regression can also achieve a satisfactory analysis upon highly nonlinear relationships between factors and response variables, while its preserving interpretability is highly valued in further analysis and experiment.

### 4.2. User Need

First of all, we must confirm whether it is necessary to provide customers with rich data resources and forms and whether this can improve the efficiency and high value of search results to a certain extent. In response, we collected back-end data from Baidu, Sougou, and Bing, sampled 200 search data users and produced the following picture:

Table1. Back-end data from Baidu, Sougou, and Bing

| Platform | Baidu | Sougou | Bing |
| --- | --- | --- | --- |
| One Search | 24 | 54 | 33 |
| Ratio–I | 12% | 27% | 16.5% |
| Twice Search | 68 | 82 | 88 |
| Ratio–II | 34% | 41% | 44% |
| Multiform Search | 108 | 64 | 79 |
| Ratio–III | 54% | 32% | 39.5% |

It can be seen from the data that only a small part of the users can find the data or information they want through a single search, and most users need a second search. We can see what they need. The proportion of users who conduct multiple search forms also means that a large user group needs multiple forms of information or data. This analysis also finds practical use-value for the application of R.

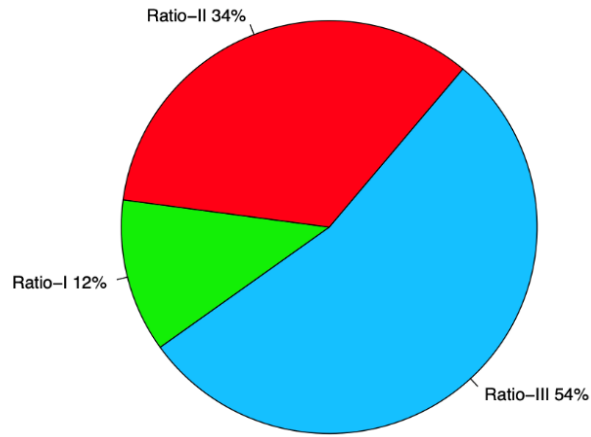

Figure 1. Baidu's user data about search time ratios

First is Baidu's user data. We can see that a search can only meet the needs of one over ten of the users. Users with need for a secondary search and compound search comprise 88% of the user community. It is essential for search engines to discover potential customer groups. They need a search engine to provide more efficient service after typing keywords, which shows information and data to meet users' needs.

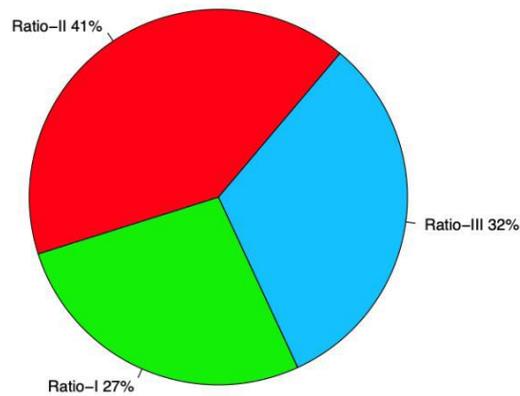

Figure 2. Sougou's user data about search time ratios

Second is the user data from Sougou. Here, we can see that 27% of the users perform a search and get the resources they need. However, there are still 32% of the users needing to search for a variety of forms. 41% of the users need to undertake a secondary search, which means that more than two-thirds of the user also has the search efficiency room for improvement.

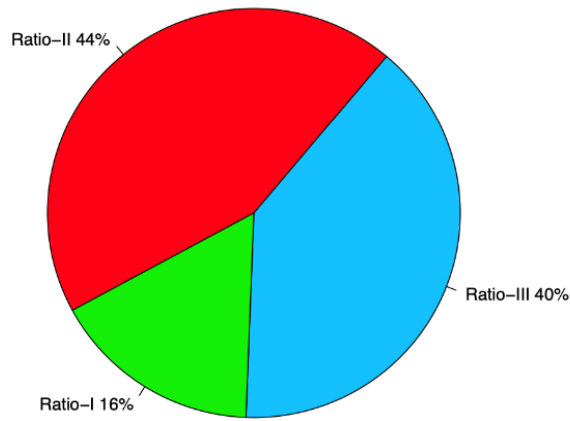

Figure 3. Bing's user data about search time ratios

Bing and Baidu, Sougou data have something in common: one time of search can only meet a few people's needs, secondary search occupies the most proportion. This kind of situation implies to search engine providers that users might abandon their search scheme and urgently need more advanced search solution to meet their new requirements.

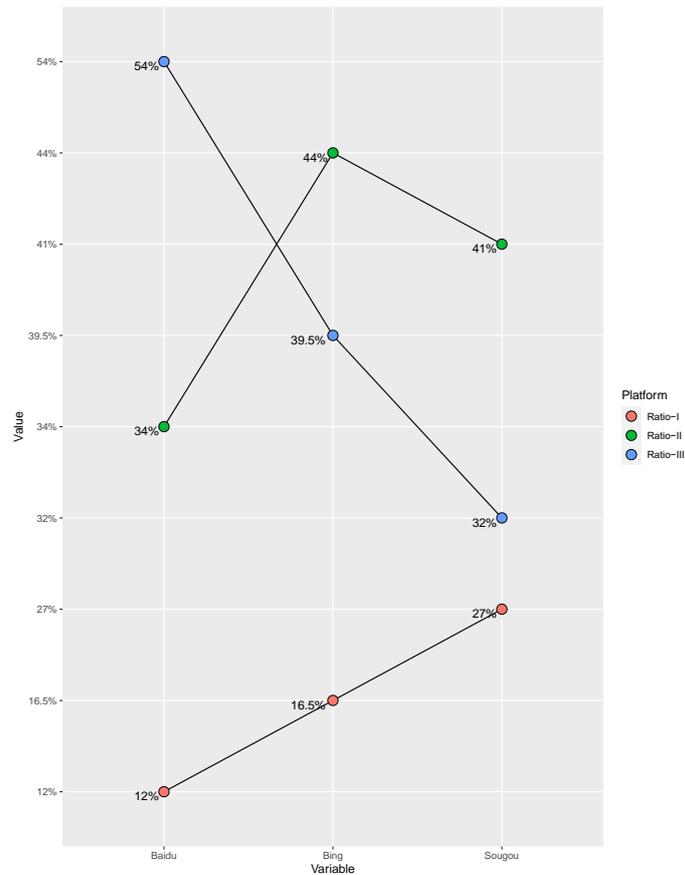

Figure 4. Line Chart of user data comparison among Baidu, Sougou, and Bing

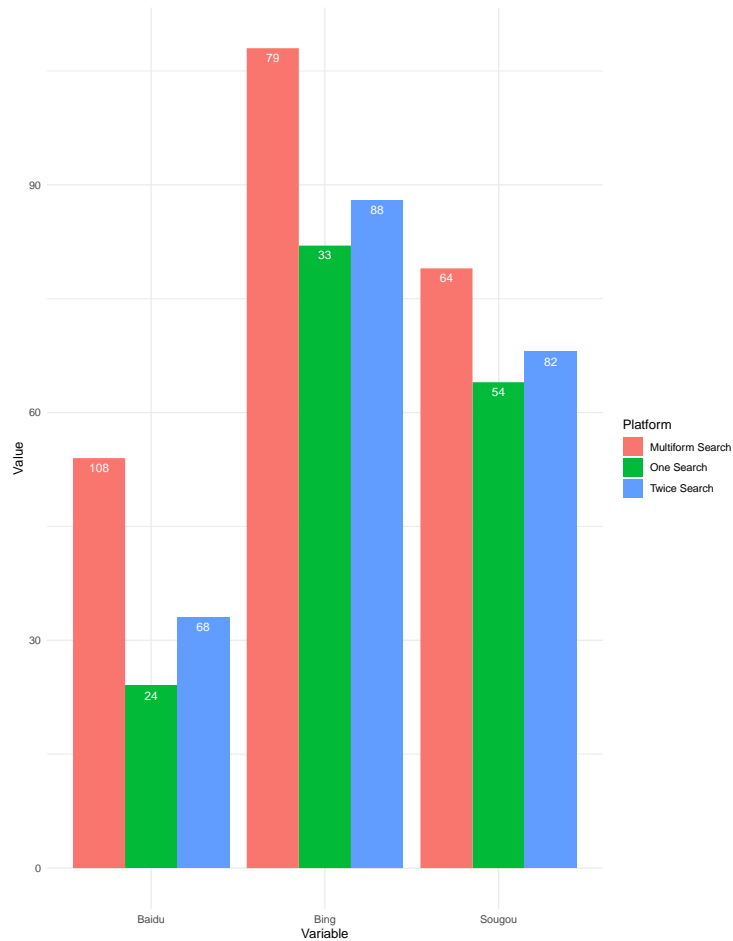

Figure 5. Histogram of user data comparison among Baidu, Sougou, and Bing

After the user enters the key search term, the qualifier is actively pushed, and the relevant qualifier is actively provided based on statistical analysis. The search user is guided to complete the final search requirements and search for satisfactory results.

### 4.3. Program Frame

A DBI-compatible interface to ODBC databases.

Depends: R ($\geq$ 3.2.0)
Imports: bit64, blob ($\geq$ 1.2.0), DBI ($\geq$ 1.0.0), hms, methods, rlang, Rcpp ($\geq$ 0.12.11)
LinkingTo: Rcpp
Suggests: covr, DBItest, magrittr, RSQLite, testthat, tibble
Published: 2020-10-27
Author: Jim Hester [aut, cre], Hadley Wickham [aut], Oliver Gjoneski [ctb] (detule), lexicalunit [cph] (nanodbc library), Google Inc. [cph] (cctz library), RStudio [cph, fnd]
Maintainer: Jim Hester <jim.hester at rstudio.com>
BugReports: https://github.com/r-dbi/odbc/issues
License: MIT + file LICENSE
URL: https://github.com/r-dbi/odbc, https://db.rstudio.com
NeedsCompilation: yes
SystemRequirements: C++11, GNU make, An ODBC3 driver manager and drivers.

Materials:       README NEWS
In views:        Databases
CRAN checks:  odbc results

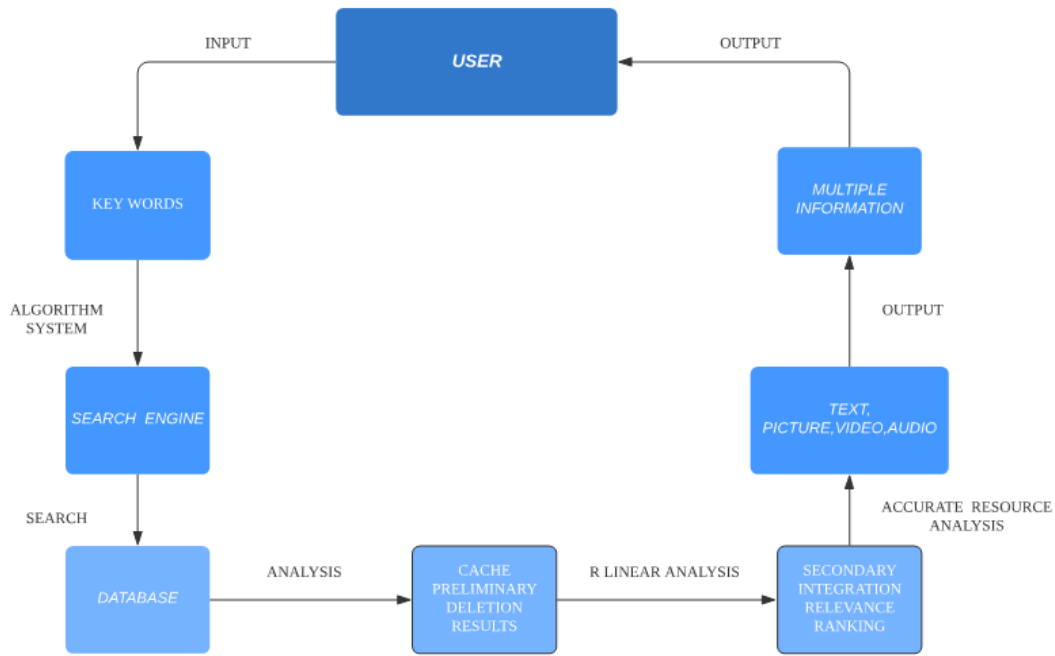

Figure 6. Program Frame for R-based Hadoop Algorithm

1. The user enters keywords and searches.
2. Search engine mobilizes Algorithm system.
3. The data of the database is read from the resource library by Hadoop.

4. Key point ①: At this time, R will perform linear analysis based on the retrieved data and find the query results that best meet the user's needs. This linear analysis is based on the user's daily usage habits after the Algorithm system is deleted. It will re-analyze the nature of keywords typed in the field of interest, self-set restrictions, and list the data with the most substantial linear relationship: the data information with the closest $R2$ to 1, based on the user's correlation has set usage habits. The DBI package provides a database interface definition for communication between R and relational database management systems. It's worth noting that some packages try to follow this interface definition (DBI-compliant) but many existing packages don't.

5. Key point ②: Afterwards, R will actively load different forms of output content, such as text, pictures, video, audio, according to the resource format. The RODBC package provides access to databases through an ODBC interface.

The RMariaDB package provides a DBI-compliant interface to MariaDB and MySQL.

The RMySQL package provides the interface to MySQL. Note that this is the legacy DBI interface to MySQL and MariaDB based on old code ported from S-PLUS. A modern MySQL client based on Rcpp is available from the RMariaDB package we listed above.

6. Display colourful forms through search engine output interface. The odbc package provides a DBI-compliant interface to drivers of Open Database Connectivity (ODBC), which is a low-level, high-performance interface that is designed specifically for relational data stores.

The RPresto package implements a DBI-compliant interface to Presto, an open source distributed SQL query engine for running interactive analytic queries against data sources of all sizes ranging from gigabytes to petabytes.
7. The user obtains the required information.
8. End of search task.

### 4.4. Experiment

Our experiment focuses on whether the search accuracy could be improved with our R-based Hadoop system. We use the same scale according to our previous back-end data from Baidu, Sougou, and Bing: "Once Search", "Twice Search", and "Multiform Search". Ideally, we hope to prioritize the increase in the ratio of "Once Search" and "Twice Search",and reduce the ratio of "Multiform Search" since this form of search means an inefficient experience for the users. With our training data, we read the database data from the resource library by Hadoop, which is a series of web links according to the entered keywords by users.

Then we mark the response variable according to the actual user behaviors. A link would be marked as "Once Search" if the user runs one search and clicks the link, "Twice Search" if the user runs two searches and clicks the link, "Multiform Search" if the user runs more than two searches or make edits, and clicks the link, "Futile Search" if the user doesn't click the link. However, our analysis will focus primarily on the first three categories of our response variable since "Futile Search" doesn't indicate a successful search in our model, but these failed attempts, with huge data, might contain information that helps improve our model accuracy.

Then we add parameters/predictors for our response variable from two parts. The first part is based on the properties of the web link, and we use the historical click rate, the relative popularity of the publisher, existence of image/audio/ external links, etc. The second part is based on usage habits from users, and we use usage frequencies of certain search engines along with personal settings, etc. After finishing the data collection and organization, we conduct near-zero-variance predictors elimination, highly correlated predictors elimination, centering and scaling of predictors, linear regression summary, and principal component analysis (PCA) to filter the most significant predictors. Then, with repeated cross-validation, we apply four machine learning models based on training data: KNN, LDA, QDA, and Multinomial logistic regression, and take a model ensemble based on the majority vote. After the training of our ensemble models, we use test data from our previous data to see if the ratios of "Once Search" and "Twice Search" increase. The followings are our test results after removing the "Futile Search" and selecting the same total size for our first three categories:

Table 2. Search Time Ratios from Baidu, Sougou, and Bing after linear regression

| Platform | Baidu | Sougou | Bing |
|---|---|---|---|
| One Search | 55 | 57 | 75 |
| Ratio−I | 27.5% | 28.5% | 37.5% |
| Twice Search | 42 | 75 | 66 |
| Ratio−II | 21% | 37.5% | 33% |
| Multiform Search | 103 | 68 | 59 |
| Ratio−III | 51.5% | 34% | 29.5% |

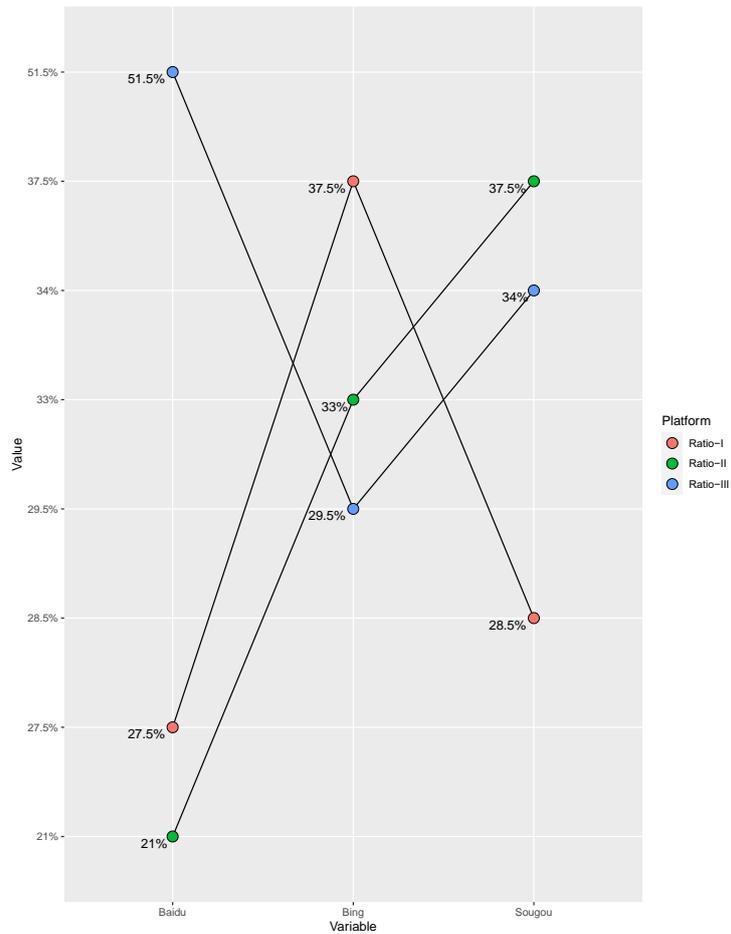

Figure 7. Line Chart of user data comparison among Baidu, Sougou, and Bing after linear regression

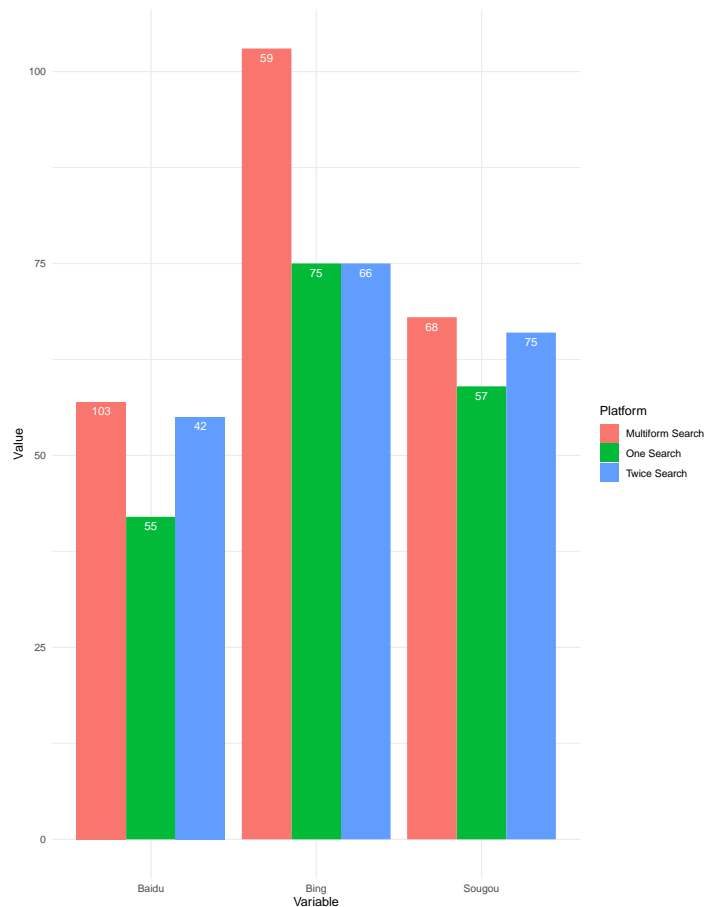

Figure 8. Histogram of user data comparison among Baidu, Sougou, and Bing after linear regression

In conclusion, theoretically, a secondary process with R linear regression model for the Hadoop database can markedly increase the ratios of "Once Search" and "Twice Search", while reducing the inefficient experience for users brought by "Multiform Searches". This is only a quite primitive attempt with R's basic machine learning functions, and we can definitely apply more complex strategies, like neural network and backward propagation, to further increase the accuracy of the algorithm to best fit the users' needs.

## 5. CONCLUSION

The role of search engines is to provide rich information and data to meet user needs. User activity is increasing, and the requirements for search engines are becoming more and more diverse. Realizing the leapfrog development of search engines and meeting users' needs for rich information resources and diverse data is the development direction of contemporary search engine suppliers.

Based on Hadoop's significant data analysis capability, different search optimization solutions can be better formulated for different users; and the system integration of R can be used as a built-in system program here. Through the analysis of information, the best selection is selected. The user's needs are highly fitted to the information flow and achieved in one step, achieving a significant leap in human-computer interaction. This procedure should be the goal of searching for users and the ultimate goal of the server: Reduce unnecessary secondary search along with multi-form search and use the most straightforward operation to achieve the most valuable information aggregation.

The current paper only proves the market value and uses the value of using R to improve search efficiency from the user's point of view. This new algorithm is achieved through the combination of Hadoop and R. With a personalized regression analysis for individual users, the search engine might achieve an optimized resources integration and significantly reduce the number of the secondary and multi-form searches. To realize this new algorithm, this article also provides a program frame for its analysis procedures. However, this proposal has not been fully verified, nor has it been tested. In this paper, R is not discussed in detail, and the cited demonstration data are not rigorous enough, the data sampling is not comprehensive, and the age and gender of users are not limited. This issue is the shortcoming of this paper, which needs further investigation.

Computer Science & Information Technology (CS & IT) 91

**AUTHOR**

I am a fourth-year student in University of California, Los Angeles. I double-major in Economics and Statistics, with a minor in Mathematics. For my internships, I used to work as a high-frequency trader in Citadel Securities, Chicago, IL, and an executive director assistant in J P Morgan, London, UK. I also worked as a research assistant in Institute of Computing Technology, Chinese Academy of Science, for Distributed Computing System, Big Data, Architecture and Machine Learning.

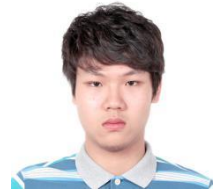